\documentclass[Physsubmission, Phys]{SciPost}
\hypersetup{
    colorlinks,
    linkcolor={red!50!black},
    citecolor={blue!50!black},
    urlcolor={blue!80!black}
}

\usepackage[bitstream-charter]{mathdesign}
\usepackage{caption}
\usepackage{subcaption}
\usepackage{slashed}

\DeclareSymbolFont{usualmathcal}{OMS}{cmsy}{m}{n}
\DeclareSymbolFontAlphabet{\mathcal}{usualmathcal}

\begin{document}

% TODO: write your article's title here.
% The article title is centered, Large boldface, and should fit in two lines
\begin{center}{\Large \textbf{
A pitfall in applying non-anticommuting $\gamma_5$ in $q\overline{q} \rightarrow ZH$ amplitudes\\
}}\end{center}

% TODO: write the author list here. Use initials + surname format.
% Separate subsequent authors by a comma, omit comma at the end of the list.
% Mark the corresponding author with a superscript *.
\begin{center}
Taushif Ahmed
% ,
% Aah B. Cee\textsuperscript{2} and
% Gee K. See\textsuperscript{3$\star$}
\end{center}

% TODO: write all affiliations here.
% Format: institute, city, country
\begin{center}
Dipartimento di Fisica and Arnold-Regge Center, Universit\`a di Torino, 
\\ and INFN, Sezione di Torino\\ Via Pietro Giuria 1, I-10125 Torino, Italy
% \\
% {\bf 2} Affiliation2
% \\
% {\bf 3} Affiliation2
\\
% TODO: provide email address of corresponding author
taushif.ahmed@unito.it
\end{center}

% \begin{center}
% \today
% \end{center}

% For convenience during refereeing (optional),
% you can turn on line numbers by uncommenting the next line:
%\linenumbers
% You should run LaTeX twice in order for the line numbers to appear.

\definecolor{palegray}{gray}{0.95}
\begin{center}
\colorbox{palegray}{
  \begin{tabular}{rr}
  \begin{minipage}{0.1\textwidth}
    \includegraphics[width=35mm]{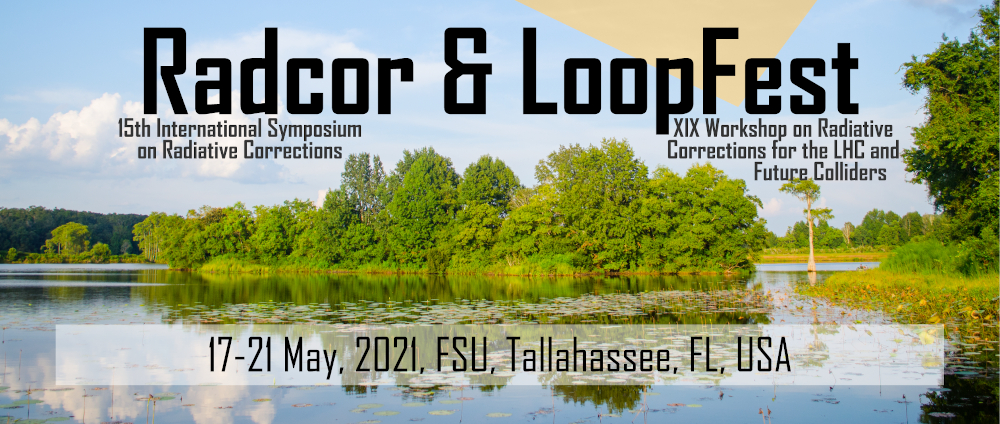}
  \end{minipage}
  &
  \begin{minipage}{0.85\textwidth}
    \begin{center}
    {\it 15th International Symposium on Radiative Corrections: \\Applications of Quantum Field Theory to Phenomenology,}\\
    {\it FSU, Tallahasse, FL, USA, 17-21 May 2021} \\
    %\doi{10.21468/SciPostPhysProc.?}\\
    \end{center}
  \end{minipage}
\end{tabular}
}
\end{center}

\section*{Abstract}
{\bf
% TODO: write your abstract here.
In computing the two-loop QCD corrections to a class of Feynman diagrams for the process $q\overline{q} \rightarrow ZH$ in Higgs effective field theory, we discover a striking phenomenon. We find the need for an additional local composite operator in the renormalised Lagrangian while employing a non-anticommuting $\gamma_5$ in dimensional regularisation. The computation using anticommuting $\gamma_5$, however, does not require any such amendment.
}

% TODO: include a table of contents (optional)
% Guideline: if your paper is longer that 6 pages, include a TOC
% To remove the TOC, simply cut the following block
\vspace{10pt}
\noindent\rule{\textwidth}{1pt}
\tableofcontents\thispagestyle{fancy}
\noindent\rule{\textwidth}{1pt}
\vspace{10pt}

%%%%%% section %%%%%%%%
\section{Introduction}
\label{sec:intro}
In exploration of the discovered scalar resonance at the Large Hadron Collider, the VH events play an important role. The recent event of the direct observation~\cite{Aaboud:2018zhk,Sirunyan:2018kst} of the Higgs boson decay to a pair of bottom quarks gets its primary contribution from the VH channel. Owing to its high phenomenological importance, there has been a lot of computations aiming to make the theoretical predictions more precise~\cite{Brein:2003wg,Brein:2011vx,Ferrera:2014lca,Ahmed:2014cla,Li:2014bfa,Catani:2014uta,Kumar:2014uwa,Campbell:2016jau,Ferrera:2017zex,Ahmed:2019udm}. In this article, we compute the two-loop QCD corrections to the production of the scalar Higgs boson in association with the neutral massive vector boson through quark annihilation in Higgs effective field theory. Through this computation, we discover a striking phenomenon. Employing non-anticommuting $\gamma_5$ in dimensional regularisation fails to generate the amplitude fulfilling the expected chiral invariance and Ward identity. The restoration of these essential properties demands amendment of a four-point local composite operator to the renormalised Lagrangian. On the other hand, applying anticommuting $\gamma_5$ gives rise to expected results satisfying these criteria. 
\\~

In the Higgs effective field theory (HEFT), where the top quark loop is integrated out by treating its mass ($m_t$) infinitely large, the interacting Lagrangian density relevant for the single scalar Higgs boson production reads as
\begin{align}
\label{eq:HEFT-Lag}
    {\cal L}_{\rm heff}= -\frac{1}{4} c_t C_{H}  \frac{H}{v} G^a_{\mu\nu} G^{a,\mu\nu}\,,
\end{align}
where $G^{a}_{\mu\nu}$, $H$, $v$ and $C_H$ denote the gluon field strength tensor, scalar Higgs boson, vacuum expectation value of the Higgs field and Wilson coefficient, respectively. We are interested in the massless QCD corrections to third order to the  production of a massive neutral vector boson, $Z$, and the $H$ through quark annihilation within the HEFT, i.e.
\begin{align}
\label{eq:scattering-process}
q(p_1)+\overline{q}(p_2) \rightarrow Z(q_1)+H(q_2)\,,
\end{align}
where $q$($\overline{q}$) denotes a quark (anti-quark). The four-momenta appearing inside the parentheses satisfy the on-shell conditions $p_1^2=p_2^2=0$ and $q_1^2=m_Z^2$, $q_2^2=m_H^2$ with $m_Z$ and $m_H$ being the mass of $Z$ and $H$ boson, respectively. 

\begin{figure}[h]
	\centering
	\begin{subfigure}[c]{.2\textwidth}
		\raisebox{-\height}{\includegraphics[width=0.8\textwidth,angle=90
		]{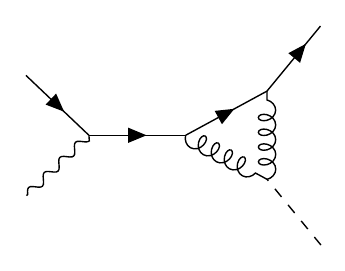}}
	\end{subfigure}
\hspace{1cm}
	\begin{subfigure}[c]{.2\textwidth}
		\raisebox{-\height}{\includegraphics[width=0.8\textwidth,angle=0]{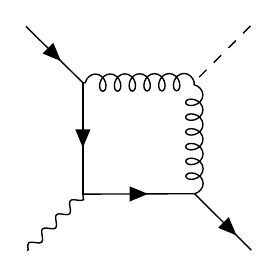}}
	\end{subfigure}
	\caption{Sample diagrams at the leading order in HEFT. The external curly and dotted lines respectively denote the $Z$ and $H$-bosons.}
	\label{fig:1L}
\end{figure}

In the original theory, the two-loop corrections proportional to top-Higgs Yukawa coupling $\lambda_t$ can be classified into two categories: class-I and class-II, depending on whether the $Z$ boson couples to the external light or the top quark loop, giving rise to different electroweak coupling factors. In \eqref{eq:HEFT-Lag}, the presence of a dimensionless parameter $c_t$ with unit value signifies that we are restricting ourselves to the class-I within HEFT, as shown in figure~\ref{fig:1L}.
%1L
In ref.~\cite{Brein:2011vx}, it was demonstrated through heavy-mass expansion to class-I, that the leading term in $1/m_t$ does not involve the effective vertex $q\overline{q}ZH$ or $q\overline{q}H$. Consequently, the leading approximation can equivalently be captured through the aforementioned effective Lagrangian \eqref{eq:HEFT-Lag} where $H$ couples to only gluons. \textit{Through this article}, we will discover that the validity of this statement depends on how we regularise the axial current, depending on whether we adopt an anticommuting or non-anticommuting $\gamma_5$, the statement holds true or fails, respectively. We start by computing the quantum loop corrections to this process using anticommuting $\gamma_5$ in the next section.

%%%%%% section %%%%%%%%
\section{Form factors employing anticommuting $\gamma_5$}
\label{eq:ACg5}
The amplitude $\mathcal{A}$ can be decomposed into vector ($\mathcal{A}_{vec}$) and axial ($\mathcal{A}_{axi}$) parts as
\begin{align} 
\label{eq:qqZHamplitude}
    \mathcal{A} &= c_t~g_{V,q}~\bar{v}(p_2) \, \mathbf{\Gamma}^{\mu}_{vec} \, u(p_1) \, \varepsilon^{*}_{\mu}(q_1) 
    \,+\, c_t~\bar{v}(p_2) \, \mathbf{\Gamma}^{\mu}_{axi} \, u(p_1) \, \varepsilon^{*}_{\mu}(q_1)\equiv c_t~g_{V,q}~{\cal A}_{vec}+ c_t~{\cal A}_{axi}\,,
\end{align}
where we have factored out the vector coupling $g_{V,q}$ between the $Z$ boson and light quark. The polarisation vector of the $Z$ boson is denoted by $\varepsilon_\mu$. By performing the Lorentz covariance decomposition of $\mathcal{A}_{vec}$ in terms of linearly independent and complete Lorentz structures in D($=4-2\epsilon$)-dimensions as
\begin{align} 
\label{eq:FFdecomp-qqZH-vec}
\mathcal{A}_{vec}^\mu &= \bar{v}(p_2) \left(
    {\cal F}_{1,vec} \slashed{q_1}  p_1^{\mu} 
    + {\cal F}_{2,vec}  \slashed{q_1}  p_2^{\mu} + {\cal F}_{3,vec}   \slashed{q_1}  q_1^{\mu}
    + {\cal F}_{4,vec}  \gamma^{\mu} \right)u(p_1) \,.
\end{align}
While performing the decomposition, we make sure the chirality is conserved along the massless quark line. From the tensorial structures present in \eqref{eq:FFdecomp-qqZH-vec}, we can construct a set of four projectors in D-dimensions which can subsequently be applied on the set of Feynman diagrams to compute the form factors (FF) ${\cal F}_{i,vec}$ order by order in perturbation theory. 

%2L
\begin{figure}[h]
	\centering
	\begin{subfigure}[c]{.2\textwidth}
		\raisebox{-\height}{\includegraphics[width=0.8\textwidth,angle=90
		]{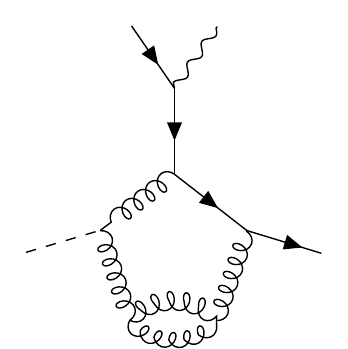}}
	\end{subfigure}
\hspace{0.4cm}
	\begin{subfigure}[c]{.2\textwidth}
		\raisebox{-\height}{\includegraphics[width=0.8\textwidth,angle=0]{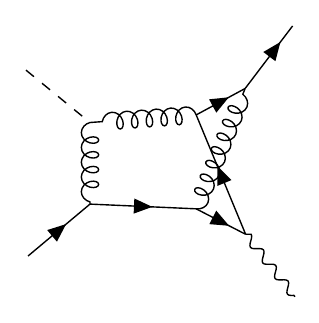}}
	\end{subfigure}
\hspace{0.6cm}
	\begin{subfigure}[c]{.2\textwidth}
		\raisebox{-\height}{\includegraphics[width=0.8\textwidth,angle=270]{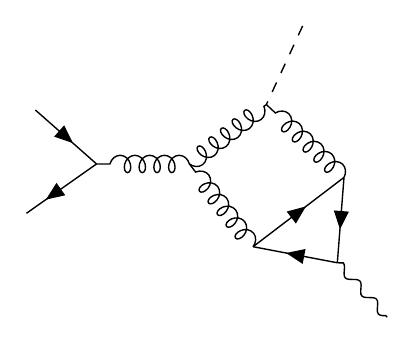}}
	\end{subfigure}
\hspace{0.4cm}
	\begin{subfigure}[c]{.2\textwidth}
		\raisebox{-\height}{\includegraphics[width=0.8\textwidth,angle=0
		]{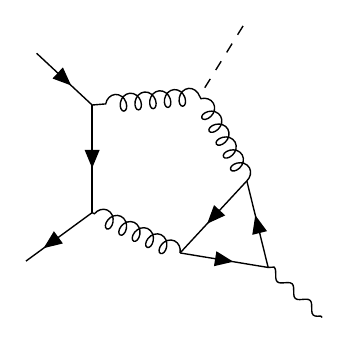}}
	\end{subfigure}
	\caption{Sample non-singlet (first two) and singlet (last two) diagrams at the next-to-leading order in HEFT}
	\label{fig:2L-ns}
\end{figure}

The axial part of the amplitude can further be categorised into non-singlet (${\cal A}_{axi(ns)}$) and singlet (${\cal A}_{axi(s)}$) components as
\begin{equation}
\label{eq:Aaxisns}
  {\cal A}_{axi} = g_{A,q}~{\cal A}_{axi(ns)} + g_{A,b}~{\cal A}_{axi(s)} \,.
\end{equation}
Through figures~\ref{fig:1L} and \ref{fig:2L-ns}, we show some sample Feynman diagrams at one- and two-loop. The singlet diagrams, featuring a closed fermionic triangle loop and exhibiting anomalous behaviour, start appearing from two-loop. By performing a Lorentz covariant decomposition of the axial part, we get
\begin{align} 
\label{eq:FFD_axi}
\mathcal{A}_{axi}^\mu &\equiv \bar{v}(p_2) \left(
    {\cal F}_{1,axi} \slashed{q_1}  p_1^{\mu} 
    + {\cal F}_{2,axi}  \slashed{q_1}  p_2^{\mu} + {\cal F}_{3,axi}   \slashed{q_1}  q_1^{\mu}
    + {\cal F}_{4,axi}  \gamma^{\mu} \right)\gamma_5 u(p_1) \,.
\end{align}
The computation of the non-anomalous part $\mathcal{A}_{axi(ns)}$, which gets translated to the computations of $\mathcal{F}_{i,axi(ns)}$, is performed using anticommuting $\gamma_5^{\rm AC}$ in D-dimensions. Due to the presence of axial anomaly, we calculate the $\mathcal{F}_{i,axi(s)}$ adopting non-anticommuting $\gamma_5$ which we discuss in the next section. By performing the strong coupling constant and operator renormalisation of \eqref{eq:HEFT-Lag}, we arrive at the ultraviolet (UV) finite set of FF to two loops which are found to contain the soft and collinear (IR) divergences, as predicted through the universal subtraction operators~\cite{Catani:1998bh}. Moreover, we find that the finite remainders ($\epsilon \rightarrow 0$) of the FF satisfy
\begin{align}
\label{eq:Ax-equal-Vec-Eff}
    {\cal F}_{i,axi(ns)}^{(l)}={\cal F}_{i,vec}^{(l)}\,, \quad i=1,2,3,4\,, \quad l=1,2\,,
\end{align}
with 
\begin{align}
\label{eq:calF-expand}
    {\cal F}_{i,va}= \sum_{l=1}^{\infty} a^{(l+1)}_s(\mu_R) {\cal F}_{i,va}^{(l)}\,, \quad va=vec,axi(ns)\,.
\end{align}
$a_s(\mu_R)\equiv\alpha_s(\mu_R)/(4\pi)$ is the strong coupling constant at the renormalisation scale $\mu_R$. The identity in \eqref{eq:Ax-equal-Vec-Eff} is a reflection of the expected chiral invariance. Our method of computation is presented in refs.~\cite{Ahmed:2019udm,Ahmed:2020kme}. We discuss the computation by employing the non-anticommuting $\gamma_5$ in the following section where we encounter a technical pitfall starting from the leading order (LO) itself.

%%%%%% section %%%%%%%%
\section{Form factors employing non-anticommuting $\gamma_5$ in HEFT}
\label{eq:NACg5}
The non-anticommuting (NAC) $\gamma_5$ can be defined in terms of Dirac's $\gamma^\mu$ matrix in dimensional regularisation as~\cite{tHooft:1972tcz,Breitenlohner:1977hr}
\begin{align}
\label{eq:gamma5}
	\gamma_5=-\frac{i}{4!}\varepsilon_{\mu\nu\rho\sigma}\gamma^{\mu}\gamma^{\nu}\gamma^{\rho}\gamma^{\sigma}\,,
\end{align}
where the Lorentz indices of the Levi-civita symbol $\varepsilon_{\mu\nu\rho\sigma}$ are treated in D dimensions~\cite{Larin:1991tj}. Owing to the usage of this definition, UV renormalisation requires an additional treatment~\cite{Larin:1993tq} to ensure the restoration of the appropriate Ward identity. However, while verifying the expected Ward identity, we discover a striking phenomenon. Although the finite remainders ($\epsilon \rightarrow 0$) of the LO vector form factors are identical to the corresponding quantities obtained using AC $\gamma_5$, the axial ones are not. In particular, we find that
\begin{align}
\label{eq:FF-NAC-AC}
&\mathcal{F}_{i,axi(ns)}^{(1),{\rm NAC}}=\mathcal{F}_{i,vec}^{(1)}\,,\quad i=1,2,3,\nonumber\\
&\mathcal{F}_{4,axi(ns)}^{(1),{\rm NAC}}\neq\mathcal{F}_{4,vec}^{(1)}\,.
\end{align}
Throughout this article, we append AC and NAC for denoting the respective $\gamma_5$ scheme. The above inequality implies the violation of the Ward identity even at the LO level
\begin{align} 
\label{eq:WI_NAC5}
q_{1,\mu} \, \mathcal{A}^{\mu,{\rm NAC}}_{axi (ns)} & \neq  0\,.
\end{align}
The restoration of the Ward identity, which is a necessity, demands an amendment term of the form
\begin{align} 
\label{eq:amendO} 
\mathcal{J}^{\mu,{\rm NAC}}
\equiv 
\mathrm{Z}_5^{h}(a_s)\, \mathrm{\mathbf{C}} \, 
\Big(\bar{v}(p_2) \,\left[\gamma^{\mu} \gamma_5\right]_{L} \, u(p_1) \Big)\, .
\end{align}
The constant factor $\mathrm{\mathbf{C}} \equiv a_s \left(-4 C_F\right) {C_H}/{v} $ collects the overall $a_s^2$ of the LO amplitude and $\left[\gamma^{\mu} \gamma_5\right]_{L}$ implies the axial vector current is renormalised according to Larin's scheme~\cite{Larin:1991tj,Larin:1993tq}. Of course, this renormalisation starts playing a role from the next-to-LO (NLO). 
\begin{figure}[htbp]
\begin{center}
\includegraphics[scale=0.30]{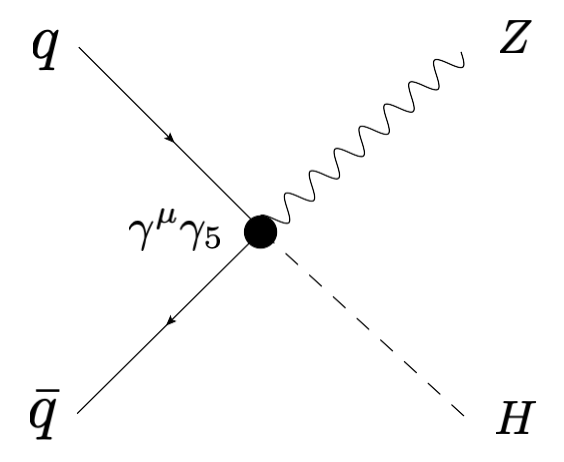}
\caption{The amendment vertex to the Lagrangian for using NAC in HEFT}
\label{dia:qqZHoperator}
\end{center}
\end{figure}
The additional renormalisation constant $\mathrm{Z}_5^{h}(a_s)=1+{\cal O}(a_s)$ that is introduced in the amendment term must be determined order by order in perturbation theory. In the upcoming subsection, we demonstrate this at the NLO. This amendment term can be visualised as a four-point local composite operator as shown in figure~\ref{dia:qqZHoperator}. With this extra term, we restore the desired properties: $\mathcal{F}^{(1),\mathrm{NAC}}_{i,axi (ns)} = \mathcal{F}_{i,vec}^{(1)}$ holds for all four FF and consequently 
$q_{1,\mu} \, \mathcal{A}^{\mu,\mathrm{NAC}}_{axi (ns)} = 0$ is also fulfilled at the LO.

The discrepancy is found to arise solely from the box diagram in figure~\ref{fig:1L}. Although each of the diagrams at the LO is individually finite, the box consists of separately diverging terms. Usage of the NAC and AC schemes of $\gamma_5$ gives rise to different D dependent coefficients of these diverging terms, which results in some non-vanishing evanescent anti-commutators upon shifting the non-anticommutating $\gamma_5$ from inside the loop to the outside. This leads to the observed discrepancy.

\subsection{UV Renormalisation of non-anomalous diagrams at the NLO}
\label{ss:UV-nonanom}
At the NLO, we have non-singlet (non-anomalous) as well as singlet (anomalous) set of diagrams, as shown in figure~\ref{fig:2L-ns}. From our experience with the LO, it is quite expected that the phenomenon shows up even at the NLO. For non-anomalous diagrams, the normal form of the Ward identity should hold for the axial current, exactly same as the vector counterpart. We find that the axial form factors renormalised according to Larin's prescription~\cite{Larin:1991tj,Larin:1993tq} exhibit the following behaviours:
\begin{enumerate}
    \item the IR pole structure, in particular, the single pole in $\epsilon$, differs from the universal prediction~\cite{Catani:1998bh},
    \item the ${\cal O}(\epsilon^0)$ term obtained after IR subtraction fails to satisfy the Ward identity.
\end{enumerate}
These shortcomings can be cured by incorporating the quantum loop corrections to the amendment vertex~\eqref{eq:amendO}. By demanding the restoration of the aforementioned properties, we get
\begin{align} 
\label{eq:amend_Z5ns} 
\mathrm{Z}_{5, ns}^{h}(a_s) &=
1 \,+\,  a_s\, \left( 
\frac{-\beta_0}{\epsilon} 
\,+\, 
\frac{107}{18} C_A   
- 7 C_F - \frac{1}{9} n_f \right)
\,+\, \mathcal{O}(a^2_s) \, ,
\end{align}
where $\beta_0=11C_A/3-2n_f/3$ is the leading coefficient of the QCD $\beta$-function, $C_A$ and $C_F$ are respectively the quadratic Casimirs in adjoint and fundamental representations, $n_f$ is the number of active light quark flavours. To summarise, the renormalised non-singlet axial amplitude $\mathcal{A}_{axi(ns)}$ in HEFT employing NAC $\gamma_5$ is given by
\begin{align}
\label{eq:renorm-ns-amp}
    {\cal A}^{\mu,\mathrm{NAC}}_{axi (ns)}(a_s) =Z^{ns}_{5,L}(a_s) Z^{ns}_{A,L}(a_s) Z_H(a_s) {\hat {\cal A}}^{\mu,\mathrm{NAC}}_{axi (ns)}({\hat a}_s) \,+\, 
    \mathcal{J}^{\mu,{\rm NAC}}_{ns} \, 
\end{align}
with the counterterm involving the amendment
\begin{align}
\label{eq:amendo-ns}
     \mathcal{J}^{\mu,{\rm NAC}}_{ns}
    &=\mathrm{Z}_{5,ns}^{h}(a_s)\, \mathrm{\mathbf{C}} \, 
    \Big(\bar{v}(p_2) \,\left[\gamma^{\mu} \gamma_5\right]_{L} \, u(p_1) \Big) \nonumber\\
    &=\mathrm{Z}_{5,ns}^{h}(a_s) Z^{ns}_{5,L}(a_s) Z^{ns}_{A,L}(a_s)\, \mathrm{\mathbf{C}} \, 
    \Big(\bar{v}(p_2) \,\gamma^{\mu} \gamma_5 \, u(p_1) \Big)\,.
\end{align}
The renormalisation constants for the non-singlet axial currents in Larin scheme are encoded through $Z^{ns}_{A,L}$ and $Z^{ns}_{5,L}$~\cite{Larin:1993tq}. We denote the operator renormalisation for \eqref{eq:HEFT-Lag} by $Z_H$. The symbol hat($~\hat{}~$) is used to indicate the bare amplitude that we get directly from Feynman diagrams. If we choose not to invoke the Larin counterterms in \eqref{eq:amendo-ns}, we could only determine the product $\mathrm{Z}_{5,ns}^{h,\mathrm{T}}(a_s) \equiv \mathrm{Z}_{5,ns}^{h}(a_s) Z^{ns}_{5,L}(a_s) Z^{ns}_{A,L}(a_s)$ as a whole.

\subsection{UV Renormalisation of anomalous diagrams}
\label{ss:UV-anom}
\begin{figure}[htbp]
\begin{center}
\includegraphics[scale=0.32]{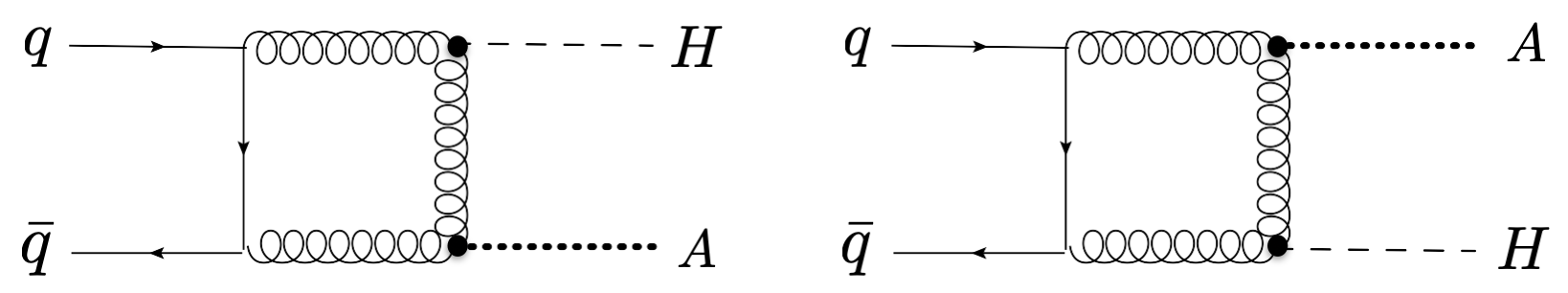}
\caption{Diagrams contributing to the right-hand side of \eqref{eq:WardIABJ}. The blobs denote the effective vertex.}
\label{dia:ABJ_RHS}
\end{center}
\end{figure}

Treatment of the singlet diagrams, as shown in figure~\ref{fig:2L-ns}, is trickier owing to the presence of ABJ anomaly~\cite{Adler:1969gk,Bell:1969ts}. The only nonzero contribution in singlet diagrams comes from the massless $b$-quark loop in $n_f=5$ flavour HEFT, that involves the coupling factor $g_{A,b} c_t$. Other generations of quarks do not contribute due to the mass degeneracy. The ABJ anomaly of the massless axial vector $b$-quark current $J_{5\mu}={\bar b}\gamma_\mu \gamma_5 b$ at the operator level reads as
\begin{align}
\label{eq:ABJ}
    \Big[\partial^{\mu} J_{5\mu}\Big]_R=a_s\frac{1}{2}\Big[G\tilde G\Big]_R \, ,
\end{align}
where $G\tilde G =-\epsilon^{\mu\nu\rho\sigma}G^a_{\mu\nu}G^a_{\rho\sigma}$. We put the subscript $R$ to signify the relation holds for renormalised operators~\cite{Larin:1993tq,Ahmed:2021spj}. The divergence of $J_{5\mu}$ gets translated to the replacement $\varepsilon_{\mu}^{*} \to  q_{1,\mu}$. At the matrix element level in momentum space, we get
\begin{equation}
 \mathcal{A}_{axi (s)}^\mu q_{1,\mu} =
  \frac{a_s}{2} \big\langle H(q_2)\big| \big[G \tilde{G}(0)\big]_R \big| q(p_1) {\overline q}(p_2) \big\rangle 
\label{eq:WardIABJ}
\end{equation}
obeying the kinematic relation $p_1+p_2 -q_2=q_1$. The above Green's function has an insertion of the composite operator $\big[G \tilde{G}(0)\big]_R$ with momentum $q^{\mu}_1$. Note that, the l.h.s. as well as the r.h.s. is individually finite. By computing the r.h.s. in perturbation theory, for which some diagrams are shown in figure~\ref{dia:ABJ_RHS},
we find that the relation \eqref{eq:WardIABJ} fails to hold true. In view of our treatment for non-anomalous diagrams at the NLO, we introduce the local composite operator
\begin{align}
\label{eq:amendo-s}
     \mathcal{J}^{\mu,{\rm NAC}}_{s}
    &=\mathrm{Z}_{5,s}^{h}(a_s)\, \mathrm{\mathbf{C}} \, 
    \Big(\bar{v}(p_2) \,\left[\gamma^{\mu} \gamma_5\right]_{L} \, u(p_1) \Big) 
\end{align}
with a new undetermined renormalisation constant $\mathrm{Z}_{5,s}^{h}(a_s)$. By demanding the finite anomaly and restoration of the anomalous Ward identity \eqref{eq:ABJ} or \eqref{eq:WardIABJ}, we obtain
\begin{align} 
\label{eq:amend_Z5s} 
\mathrm{Z}_{5, s}^{h}(a_s) &=
1 \,+\,  a_s\, \left( 
-\frac{3}{2}\frac{1}{\epsilon} 
-\frac{3}{4} \right) 
\,+\, \mathcal{O}(a^2_s) \, .
\end{align}
In the process of renormalising the matrix elements on the r.h.s. of \eqref{eq:WardIABJ},  we require the following mixed counterterm
\begin{align}
\label{eq:amend_ZGJ}
\mathrm{Z}_{GJ}^{h}(a_s)\,  
\big(\bar{v}(p_2) \,\gamma_{\mu} \gamma_5 \, u(p_1) \, q^{\mu}_{1} \big)      
\end{align}
where $\mathrm{Z}_{GJ}^{h}(a_s) = a_s\, \left(\frac{24 } {\epsilon} C_F\right) + \mathcal{O}(a_s^2)$ is determined to minimally subtract all the poles in figure~\ref{dia:ABJ_RHS}.

To summarise, the renormalised effective Lagrangian with a non-anticommuting $\gamma_5$ for computing the class-I diagrams of $q\bar{q} \rightarrow ZH$ in HEFT reads as
\begin{align}
\label{eq:Rlagrangian}
\mathcal{L}_{R} = \Big[\mathcal{L}_{c}+\mathcal{L}_{\mathrm{heff}}\Big]_{R} + 
\kappa~\mathrm{Z}_5^{h}(a_s)\, \mathrm{\mathbf{C}} \, \Big(\bar{q}_R(x) \,\left[\gamma^{\mu} \gamma_5\right]_{L} \, q_R(x)\Big) Z_{\mu}(x)\, H(x)\,,
\end{align}
where $\kappa_{ns} = c_t g_{A,q}$ and  $\kappa_{s} = c_t g_{A,b}$. The first term is the renormalised Lagrangian without the new amendment, as captured through the second term. The latter does not arise while computing the QCD corrections to non-anomalous set of diagrams employing the anticommuting $\gamma_5$ scheme.

\subsection{Form factors in exact theory without heavy-top limit}
\label{ss:FF-exact}
To make our understanding more concrete, we intend to examine whether the need of the additional local composite operator persists in the full theory with $n_f=6$ while using non-anticommuting $\gamma_5$. 
\begin{figure}[htbp]
\begin{center}
\includegraphics[scale=0.30]{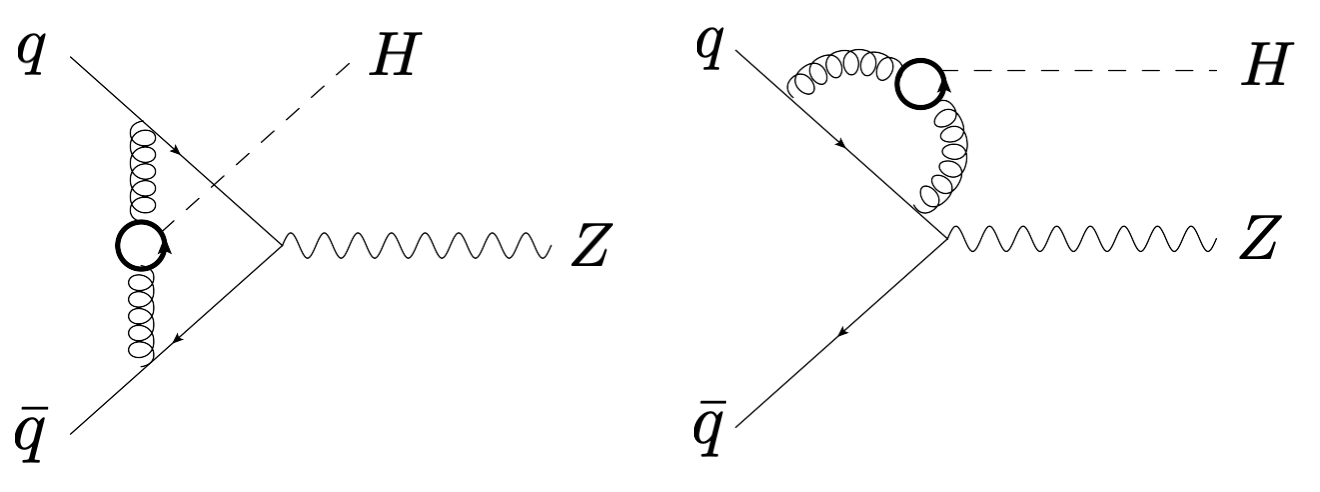}
\caption{Sample diagrams of the two-loop class-I which are proportional to
 $\lambda_t$. The thick solid lines denote the massive top quark.}
\label{dia:LO_full}
\end{center}
\end{figure}
We have 6 Feynman diagrams at two-loop with a finite top mass, with samples shown in figure~\ref{dia:LO_full}. We generate the integrand using GoSam~\cite{Cullen:2014yla,Jones:2016bci} and apply the integration-by-parts~\cite{Tkachov:1981wb,Chetyrkin:1981qh} relations employing Kira~\cite{Maierhofer:2018gpa} to express the bare amplitude in terms of 55 master integrals. Using pySecdec~\cite{Borowka:2017idc}, we evaluate the integrals numerically at one chosen kinematic point as this is sufficient for our purpose. We find that the vector and axial form factors of the two-loop class-I diagrams indeed are identical to each other without the need of any amendment to the Lagrangian. In particular, the 4-th vector and axial FF agree to the fourth significant digit. Therefore, although we need to introduce an effective four-point vertex in HEFT while working with a non-anticommuting $\gamma_5$, we do not need to do so in the exact theory with $n_f=6$.

\section{Conclusion}
Our exploration with the class-I diagrams in HEFT ($n_f=5$) shows that the correct results obeying the chiral invariance as well as the appropriate Ward identity can be obtained for the vector and non-anomalous axial amplitude with anticommuting $\gamma_5$. However, for non-anticommuting $\gamma_5$, we need an additional four-point effective vertex  $q\gamma_\mu \gamma_5 \bar{q}Z^{\mu}H$. In the exact theory with $n_f=6$ flavours, we do not need any additional composite operator of this kind despite using non-anticommuting $\gamma_5$. This observation strengthens the common lore that a conclusion obtained employing anticommuting $\gamma_5$ does not necessarily holds for non-anticommuting $\gamma_5$ in dimensional regularisation. For which class of processes the need of this kind of effective operator arises is an interesting arena that we intend to explore in months.

\section*{Acknowledgements}
The author, together with Werner Bernreuther, Long Chen, and Micha\l{} Czakon thank the organisers of RADCOR-LoopFest 2021.

% TODO: include funding information
\paragraph{Funding information}
The author received funding from the European
Research Council (ERC) under the European Union’s Horizon 2020 research and innovation programme \textit{High precision multi-jet dynamics at the LHC} (ERC Condsolidator grant agreement No 772009).

% SECOND OPTION:
% Use your bibtex library
%\bibliographystyle{SciPost_bibstyle} % Include this style file here only if you are not using our template
\bibliography{SciPost_LaTeX_Template.bib}

\nolinenumbers

\end{document}